%% file: template.tex
\documentclass{article}
\author{Jeroen Ooms}

\usepackage{url}
\usepackage{fullpage}
\usepackage{xspace}
\usepackage{booktabs}
\usepackage{enumitem}
\usepackage[hidelinks]{hyperref}
\usepackage[round]{natbib}
\usepackage{fancyvrb}
\usepackage[toc,page]{appendix}
\usepackage{breakurl}

\usepackage{float}
\restylefloat{table}

\usepackage[utf8]{inputenc}

\usepackage{pmboxdraw}

\usepackage{parskip}

\usepackage{setspace}
\newcommand{\MCMC}{\texttt{MCMC}\xspace}
\newcommand{\URL}{\texttt{URL}\xspace}
\newcommand{\DSL}{\texttt{DSL}\xspace}
\newcommand{\REST}{\texttt{REST}\xspace}
\newcommand{\HTML}{\texttt{HTML}\xspace}
\newcommand{\Linux}{\texttt{Linux}\xspace}
\newcommand{\GET}{\texttt{GET}\xspace}
\newcommand{\POST}{\texttt{POST}\xspace}
\newcommand{\JSON}{\texttt{JSON}\xspace}
\newcommand{\SQL}{\texttt{SQL}\xspace}
\newcommand{\CGI}{\texttt{CGI}\xspace}
\newcommand{\R}{\texttt{R}\xspace}
\newcommand{\Julia}{\texttt{Julia}\xspace}
\newcommand{\SPSS}{\texttt{SPSS}\xspace}
\newcommand{\JavaScript}{\texttt{JavaScript}\xspace}

\newcommand{\HTTP}{\texttt{HTTP}\xspace}
\newcommand{\TCP}{\texttt{TCP}\xspace}
\newcommand{\IP}{\texttt{IP}\xspace}
\newcommand{\GUI}{\texttt{GUI}\xspace}
\newcommand{\CLI}{\texttt{CLI}\xspace}
\newcommand{\UI}{\texttt{UI}\xspace}
\newcommand{\API}{\texttt{API}\xspace}
\newcommand{\RPC}{\texttt{RPC}\xspace}

\newcommand{\OpenCPU}{\texttt{OpenCPU}\xspace}
\newcommand{\Latex}{\LaTeX \xspace}

\newcommand{\maintitle}[1]{
  \title{#1}
  \maketitle
}

\begin{document}
\input{content}

%references
\bibliographystyle{plainnat}
\bibliography{thesis}

%end
\end{document}

%% file: content.tex
% !TEX root = template.tex

\maintitle{The OpenCPU System: Towards a Universal Interface for Scientific Computing through Separation of Concerns}

\begin{abstract}
Applications integrating analysis components require a programmable interface which defines statistical operations independently of any programming language. By separating concerns of scientific computing from application and implementation details we can derive an interoperable \API for data analysis. But what exactly are the concerns of scientific computing? To answer this question, the paper starts with an exploration of the purpose, problems, characteristics, struggles, culture, and community of this unique branch of computing. By mapping out the domain logic, we try to unveil the fundamental principles and concepts behind statistical software. Along the way we highlight important problems and bottlenecks that need to be addressed by the system in order to facilitate reliable and scalable analysis units. Finally, the \OpenCPU software is introduced as an example implementation that builds on \HTTP and \R to expose a simple, abstracted interface for scientific computing.
\end{abstract}

% !TEX root = main.tex
\section{Introduction}

Methods for scientific computing are traditionally implemented in specialized software packages assisting the statistician in all facets of the data analysis process. A single product typically includes a wealth of functionality to interactively manage, explore and analyze data, and often much more. Products such as \R or \texttt{STATA} are optimized for use via a command line interface (\CLI) whereas others such as \SPSS focus mainly on the graphical user interface (\GUI). However, increasingly many users and organizations wish to integrate statistical computing into third party software. Rather than working in a specialized statistical environment, methods to analyze and visualize data get incorporated into pipelines, web applications and big data infrastructures. This way of doing data analysis requires a different approach to statistical software which emphasizes interoperability and programmable interfaces rather than \UI interaction. Throughout the paper we refer to this approach to statistical software as \emph{embedded scientific computing}.

Early pioneering work in this area was done by \cite{lang2000omegahat} and \cite{chambers1998distributed} who developed an environment for integration of statistical software in \texttt{Java} based on the \texttt{CORBA} standard \citep{henning2006rise}. Recent work in embedded scientific computing has mostly aimed at low-level tools for directly connecting statistical software to general purpose environments. For \R, bindings and bridges are available to execute an \R script or process from inside all popular languages. For example, \texttt{JRI} \citep{rjava}, \texttt{RInside} \citep{eddelbuettel2011rcpp}, \texttt{rpy2} \citep{gautier2008rpy2} or \texttt{RinRuby} \citep{dahl2008rinruby} can be used to call \R from respectively \texttt{Java}, \texttt{C++}, \texttt{Python} or \texttt{Ruby}. \cite{heiberger2009r} provide a set of tools to run \R code from \texttt{DCOM} clients on Windows, mostly to support calling \R in Microsoft Excel. The \texttt{rApache} module (\texttt{mod\_R}) makes it possible to execute R scripts from the \texttt{Apache2} web server \citep{horner2013rapache}. Similarly, the \texttt{littler} program provides hash-bang capability for \R, as well as simple command-line and piping use on \texttt{UNIX} \citep{littler}. Finally, \texttt{Rserve} is \texttt{TCP/IP} server which provides low level access to an \R process over a socket \citep{urbanek2013rserve}. 

Even though these language-bridging tools have been available for several years, they have not been able to facilitate the big breakthrough of \R as a ubiquitous statistical engine. Given the enormous demand for analysis and visualization these days, the adoption of \R for embedded scientific computing is actually quite underwhelming. In my experience, the primary cause for the limited success is that these bridges are hard to implement, do not scale very well, and leave the most challenging problems unresolved. Substantial plumbing and expertise of \R internals is required for building actual applications on these tools. Clients are supposed to generate and push \R syntax, make sense of \R's internal \texttt{C} structures and write their own framework for managing requests, graphics, security, data interchange, etc. Thereby, scientific computing gets intermingled with other parts of the system resulting in highly coupled software which is complex and often unreliable. High coupling is also problematic from a human point of view. Building a web application with for example \texttt{Rserve} requires a web developer that is also an expert in \R, \texttt{Java} and \texttt{Rserve}. Because \R is a very domain specific language, this combination of skills is very rare and expensive.

\subsection{Separation of concerns}

%A helpful analogy for statisticians might be provided by \texttt{SQL} servers. \texttt{SQL} separates the management of relational data from other components through an interoperable language. 
What is needed to scale up embedded scientific computing is a system that decouples data analysis from other system components in such a way that applications can integrate statistical methods without detailed understanding of \R or statistics. Component based software engineering advocates the design principle of \emph{separation of concerns} \citep{heineman2001component}, which states that a computer program is split up into distinct pieces that each encapsulate a \emph{logical} set of functionality behind a well-defined interface. This allows for independent development of various components by different people with different background and expertise. Separation of concerns is fundamental to the functional programming paradigm \citep{reade1989elements} as well as the design of service oriented architectures on distributed information systems such as the internet \citep{fielding2000architectural}. The principle lies at the heart of this research and holds the key to advancing embedded scientific computing.
%Concerns within an architecture are generally broader defined than within the context of a language, but the same rules and benefits hold. 

In order to develop a system that separates concerns of scientific computing from other parts of the system, we need to ask ourselves: what are the concerns of scientific computing? This question does not have a straightforward answer. Over the years, statistical software has gotten highly convoluted by the inclusion of complementary tools that are useful but not necessarily an integral part of computing. Separation of concerns requires us to extract the core logic and divorce it from all other apparatus. We need to form a conceptual model of data analysis that is independent of any particular application or implementation. Therefore, rather than discussing technical problems, this paper focuses entirely on studying the domain logic of the discipline along the lines of \cite{evans2004domain}. By exploring the concepts, problems, and practices of the field we try to unveil the fundamental principles behind statistical software. Along the way we highlight important problems and bottlenecks that require further attention in order to facilitate reliable and scalable analysis modules.

The end goal of this paper is to work towards an interface definition for embedded scientific computing. An interface is the embodiment of separation of concerns and serves as a contract that formalizes the boundary across which separate components exchange information. The interface definition describes the concepts and operations that components agree upon to cooperate and how the communication is arranged. In the interface we specify the functionality that a server has to implement, which parts of the interaction are fixed and which choices are specifically left at the discretion of the implementation. Ideally the specification should provide sufficient structure to develop clients and server components for scientific computing while minimizing limitations on how these can be implemented. An interface that carefully isolates components along the lines of domain logic allows developers to focus on their expertise using their tools of choice. It gives clients a universal point of interaction to integrate statistical programs without understanding the actual computing, and allows statisticians to implement their methods for use in applications without knowing specifics about the application layer.

%- think in analogy of SQL describing relational data management

\subsection{The OpenCPU system}

The \OpenCPU system is an example that illustrates what an abstracted interface to scientific computing could look like. \OpenCPU defines an \HTTP \API that builds on \emph{The R Project for Statistical Computing}, for short: \R \citep{R}. The \R language is the obvious candidate for a first implementation of this kind. It is currently the most popular statistical software package and considered by many statisticians as the de facto standard of data analysis. The huge \R community provides both the tools and use-cases needed to develop and experiment with this new approach to scientific computing. It is fair to say that currently only \R has the required scale and foundations to really put our ideas to the test. However, although the research and \OpenCPU system are colored by and tailored to the way things work in \R, the approach should generalize quite naturally to other computational back-ends. The \API is designed to describe general logic of data analysis rather than that of a particular language. The main role of the software is to put this new approach into practice and get firsthand experience with the problems and opportunities in this unexplored field. 

As part of the research, two \OpenCPU server implementations were developed. The \R package \texttt{opencpu} uses the \texttt{httpuv} web server \citep{httpuv} to implement a \emph{single-user server} which runs within an interactive \R session on any platform. The \emph{cloud server} on the other hand is a multi-user implementation based on \texttt{Ubuntu Linux} and \texttt{rApache}. The latter yields much better performance and has advanced security and configuration options, but requires a dedicated \Linux server. Another major difference between these implementations is how they handle concurrency. Because \R is single threaded, \texttt{httpuv} handles only a single request at a time. Additional incoming requests are automatically queued and executed in succession using the same process. The cloud server on the other hand takes advantage of multi-processing in the \texttt{Apache2} web server to handle concurrency. This implementation uses forks of the \R process to serve concurrent requests immediately with little performance overhead. The differences between the cloud server and single user server are invisible the client. The \API provides a standard interface to either implementation and other than varying performance, applications will behave the same regardless of which server is used. This already hints at the benefits of a well defined interface.
%The remainder of this paper focuses entirely on the logic of the system and does not treat implementation details in any further detail, apart from some details that are helpful to illustrate and exemplify the interface.

\subsection{History of OpenCPU}

The \OpenCPU system builds on several years of work dating back to 2009. The software evolved through many iterations of trial and error by which we identified the main concerns and learned what works in practice. Initial inspirations were drawn from recurring problems in developing \R web applications with \texttt{rApache}, including \cite{van2009stage}. Accumulated experiences from these projects shaped a vision on what is involved with embedded scientific computing. After a year of internal development, the first public beta of \OpenCPU appeared in August 2011. This version was picked up by early adopters in both industry and academia, some of which are still in production today. The problems and suggestions generated from early versions were a great source of feedback and revealed some fundamental problems. At the same time exciting developments were going on in the \R community, in particular the rise of the \texttt{RStudio IDE} and introduction of powerful new \R packages \texttt{knitr}, \texttt{evaluate} and \texttt{httpuv}. After a redesign of the \API and a complete rewrite of the code, \texttt{OpenCPU 1.0} was released in August 2013. By making better use of native features in \HTTP, this version is more simple, flexible, and extensible than before. Subsequent releases within the \texttt{1.x} series have introduced additional server configurations and optimizations without major changes to the \API.

% !TEX root = main.tex

\section{Practices and domain logic of scientific computing}

%The field and software of data analysis are unlike thos in any other domain. Statistical computing has a long history and many intrinsic properties and peculiarities that are relevent when embedding statistical methods into systems. It is my impression that lack of understanding both within and outside the statistics community about what distinguishes scientific computing from other branches of computer science underlies many of the technical problems. 
%At the same time there are enourmous opportunities to facilitate better science and data analysis using modern technogy that might be overlooked due to a disconnect between statisticians and software engineers. 
%The purpose of this section is to provide a high-level introduction to problems, practices and logic of scientific computing.

This section provides a helicopter view of the practices and logic of scientific computing that are most relevant in the context of this research. The reader should get a sense of what is involved with scientific computing, what makes data analysis unique, and why the software landscape is dominated by domain specific languages (\texttt{DSL}). The topics are chosen and presented somewhat subjectively based on my experiences in this field. They are not intended to be exhaustive or exclusive, but rather identify the most important concerns for developing embedded analysis components. 
%Moreover this material provides arguments for deciding if a concern is fundamental to the conceptual model and should be captured by the interface, or is better left at the discretion of the implementation.

\subsection{It starts with data}

The role and shape of \emph{data} is the main characteristic that distinguishes scientific computing. In most general purpose programming languages, \emph{data structures} are instances of classes with well-defined fields and methods. Similarly, databases use schemas or table definitions to enforce the structure of data. This ensures that a table returned by a given \SQL query always contains exactly the same structure with the requested fields; the only varying property between several executions of a query is the number of returned rows. Also the time needed for the database to process the request usually depends only on the amount of records in the database.
Strictly defined structures make it possible to write code implementing all required operations in advance without knowing the actual \emph{content} of the data. It also creates a clear separation between developers and users. Most applications do not give users direct access to raw data. Developers focus in implementing code and designing data structures, whereas users merely get to execute a limited set of operations.

This paradigm does not work for scientific computing. Developers of statistical software have relatively little control over the structure, content, and quality of the data. Data analysis starts with the user supplying a dataset, which is rarely pretty. Real world data come in all shapes and formats. They are messy, have inconsistent structures, and invisible numeric properties. Therefore statistical programming languages define data structures relatively loosely and instead implement a rich lexicon for interactively manipulating and testing the data. Unlike software operating on well-defined data structures, it is nearly impossible to write code that accounts for any scenario and will work for every possible dataset. Many functions are not applicable to every instance of a particular class, or might behave differently based on dynamic properties such as size or dimensionality. For these reasons there is also less clear of a separation between developers and users in scientific computing. The data analysis process involves simultaneously debugging of code and data where the user iterates back and forth between manipulating and analyzing the data. Implementations of statistical methods tend to be very flexible with many parameters and settings to specify behavior for the broad range of possible data. And still the user might have to go through many steps of cleaning and reshaping to give data the appropriate structure and properties to perform a particular analysis. 

Informal operations and loosely defined data structures are typical characteristics of scientific computing. They give a lot of freedom to implement powerful and flexible tools for data analysis, but complicate interfacing of statistical methods. Embedded systems require a degree of type-safety, predictability, and consistency to facilitate reliable \texttt{I/O} between components. These features are native to databases or many object oriented languages, but require substantial effort for statistical software.

\subsection{Functional programming}

Many different programming languages and styles exists, each with their own strengths and limitations. Scientific computing languages typically use a functional style of programming, where methods take a role and notation similar to \emph{functions} in mathematics. This has obvious benefits for numerical computing. Because equations are typically written as $y = f(g(x))$ (rather than $y = x.g().f()$ notation), a functional syntax results in intuitive code for implementing algorithms. 

Most popular general purpose languages take a more imperative and object oriented approach. In many ways, object-oriented programming can be considered the opposite of functional programming \citep{pythonfunctional}. Here methods are invoked on an object and modify the \emph{state} of this particular object. Object-oriented languages typically implement inheritance of fields and methods based on object classes or prototypes. Many software engineers prefer this style of programming because it is more powerful to handle complex data structures. The success of object oriented languages has also influenced scientific computing, resulting in multi-paradigm systems. Languages such as \Julia and \R use multiple dispatch to dynamically assign function calls to a particular function based on the type of arguments. This brings certain object oriented benefits to functional languages, but also complicates scoping and inheritance. 

A comparative review on programming styles is beyond the scope of this research. But what is relevant to us is how conflicting paradigms affect interfacing of analysis components. In the context of web services, the \emph{Representational State Transfer} style (for short: \REST) described by \cite{fielding2000architectural} is very popular among web developers. A \emph{restful} \API maps every \URL to a \emph{resource} and \HTTP requests are used to modify the \emph{state} of a resource, which results in a simple and elegant \API. Unfortunately, \REST does not map very naturally to the functional paradigm of statistical software. Languages where functions are first class citizens suggest more \RPC flavored interfaces, which according to Fielding are by definition not restful \citep{fielding2008rest}. This does not mean that such a component is incompatible with other pieces. As long as components honor the rules of the protocol (i.e. \HTTP) they will work together. However, conflicting programming styles can be a source of friction for embedded scientific computing. Strongly object-oriented frameworks or developers might require some additional effort to get comfortable with components implementing a more functional paradigm.

\subsection{Graphics}

Another somewhat domain specific feature of scientific computing is native support for graphics. Most statistical software packages include programs to draw plots and charts in some form or another. In contrast to data and functions which are language objects, the graphics device is considered a separate output stream. Drawings on the canvas are implemented as a side effect rather than a return value of function calls. This works a bit similar to manipulating document object model (\texttt{DOM}) elements in a browser using \JavaScript. In most interactive statistical software, graphics appear to the user in a new window. The state of the graphics device cannot easily be stored or serialized as is the case for functions and objects. We can export an \emph{image} of the graphics device to a file using \texttt{png}, \texttt{svg} or \texttt{pdf} format, but this image is merely a snapshot. It does not contain the actual state of the device cannot be reloaded for further manipulation.

First class citizenship of graphics is an important concern of interfacing scientific computing. Yet output containing both data and graphics makes the design of a general purpose \API more difficult. The system needs to capture the return value as well as graphical side effects of a remote function call. Furthermore the interface should allow for generating graphics without imposing restrictions on the format or formatting parameters. Users want to utilize a simple bitmap format such as \texttt{png} for previewing a graphic, but have the option to export the same graphic to a high quality vector based format such as \texttt{pdf} for publication. Because statistical computation is expensive and non-deterministic, the graphic cannot simply reconstructed from scratch only to retrieve it in another format. Hence the \API needs to incorporate the notion of a graphics device in a way independent of the imaging format. 

\subsection{Numeric properties and missing values}

It was already mentioned how loosely defined data structures in scientific computing can impede type safety of data \texttt{I/O} in analysis components. In addition, statistical methods can choke on the actual content of data as well. Sometimes problematic data can easily be spotted, but often it is nearly impossible to detect these ahead of time. Applying statistical procedures to these data will then result in errors, even though the code and structure of the data are perfectly fine. These problems frequently arise for statistical models that build on matrix decompositions which require the data to follow certain numeric properties. The \emph{rank} of a matrix is one such property which measures the nondegenerateness of the system of linear equations. When a matrix $A$ is rank deficient, the equation $Ax=b$ does not have a solution when $b$ does not lie in the range of $A$. Attempting to solve this equation will eventually lead to division by zero. Accounting for such cases of time is nearly impossible because numeric properties are invisible until they are actually calculated. But perhaps just as difficult is explaining the user or software engineer that these errors are not a bug, and can not be fixed. The procedure just does not work for this particular dataset.

Another case of problematic data is presented by \emph{missing} values. Missingness in statistics means that the value of a field is unknown. Missing data should not be confused with no data or \texttt{null}. Missing values are often \emph{non ignorable}, meaning that the missingness itself is information that needs to be accounted for in the modeling. A standard textbook example is when we perform a survey asking people about their salary. Because some people might refuse to provide this information, the data contains missing values. This missingness is probably \emph{not completely at random}: respondents with high salaries might be more reluctant to provide this information than respondents with a median salary. If we calculate the mean salary from our data ignoring the missing values, the estimate is likely biased. To obtain a more accurate estimate of the average salary, missing values need to be incorporated in the estimation using a more sophisticated model. 

Statistical programming languages can define several types of missing or non-finite values such as \texttt{NA}, \texttt{NaN} or \texttt{Inf}. These are usually implemented as special primitives, which is one of the benefits of using a \DSL. Functions in statistical software have built-in procedures and options to specify how to handle missing values encountered in the data. However, the notion of missingness is foreign to most languages and software outside of scientific computing. They are a typical domain-specific phenomenon that can cause technical problems in data exchange with other systems. And like numeric properties, the concept of values containing no actual value is likely to cause confusion among developers or users with limited experience in data analysis. Yet failure to properly incorporate missing values in the data can easily lead to errors or incorrect results, as the example above illustrated.

\subsection{Non deterministic and unpredictable behavior}

Most software applications are expected to produce consistent output in a timely manner, unless something is very wrong. This does not generally hold for scientific computing. The previous section explained how problematic data can cause exceptions or unexpected results. But many analysis methods are actually non-deterministic or unpredictable by nature.

Statistical algorithms often repeat some calculation until a particular convergence criterion is reached. Starting values and minor fluctuations in the data can have snowball effect on the course of the algorithm. Therefore several runs can result in wildly varying outcomes and completion times. Moreover, convergence might not be guaranteed: unfortunate input can get a process stuck in a local minimum or send it off into the wrong direction. Predicting and controlling for such scenarios a-priori in the implementation is very difficult. Monte Carlo techniques are even less predictable because they are specifically designed to behave randomly. For example, \MCMC methods use a Markov-Chain to simulate random walk through a (high-dimensional) space such as a multivariate probability density. These methods are a powerful tool for simulation studies and numerical integration in Bayesian analysis. Each execution of the random walk yields different outcomes, but under general conditions the process will converge to the value of interest. However, due to randomness it is possible that some of the runs or chains get stuck and need to be terminated or disregarded.

Unpredictability of statistical methods underlies many technical problems for embedded scientific computing that are not present when interacting with a database. This can sometimes surprise software engineers expecting deterministic behavior. Statistical methods are rarely absolutely guaranteed to be successful for arbitrary data. Assuming that a procedure will always return timely and consistently because it did so with testing data is very dangerous. In a console, the user can easily intervene or recover, and retry with different options or starting values. For embedded modules, unpredictability needs to be accounted for in the design of the system. At a very minimum, the system should be able to detect and terminate a process that has not completed when some timeout is reached. But preferably we need a layer or meta functionality to control and monitor executions, either manually or automatically. 

\subsection{Managing experimental software}

In scientific computing, we usually need to work with inventive, volatile, and experimental software. This is a big cultural difference with many general purpose languages such as \texttt{python}, \texttt{Java}, \texttt{C++} or \texttt{JavaScript}. The latter communities include professional organizations and engineers committed to implementing and maintaining production quality libraries. Most authors of open source statistical software do not have the expertise and resources to meet such standards. Contributed code in languages like \texttt{R} was often written by academics or students to accompany a scientific article proposing novel models, algorithms, or programming techniques. The script or package serves as an illustration of the presented ideas, but needs needs to be tweaked and tailored to fit a particular problem or dataset. The quality of such contributions varies a lot, no active support or maintenance should be expected from the authors. Furthermore, package updates can sometimes introduce radical changes based on new insights. 

Because traditional data analysis does not really have a notion of production, this has never been a major problem. The emphasis in statistical software has always been on innovation rather than continuity. Experimental code is usually good enough for interactive data analysis where it suffices to manually make a script or package work for the dataset at hand.
Authors of statistical software tend to assume that the user will spend some effort to manage dependencies and debug the code. However, integrated components require a greater degree of reliability and continuity which introduces a source of technical and cultural friction for embedded scientific computing. 
%This will likely not change overnight. Proposals to make statistical software play nicer with other software have been received with little understanding and sympathy within the community. 
This makes the ability to manage unstable software, facilitate rapid change, sandbox modules, and manage failure important concerns of embedded scientific computing.

\subsection{Interactivity and error handling}

In general purpose languages, run-time errors are typically caused by a bug or some sort of system failure. Exceptions are only raised when the software can not recover and usually result in termination of the process. Error messages contain information such as calling stacks to help the programmer discover where in the code a problem occurred. Software engineers go through great trouble to prevent potential problems ahead of time using smart compilers, unit tests, automatic code analysis, and continuous integration. Errors that do arise during production are usually not displayed to the user, but rather the administrator is notified that the system urgently needs attention. The user gets to see an apology at best.

In scientific computing, errors play a very different role. As a consequence of some of the characteristics discussed earlier, interactive debugging is a natural part of the user experience. Errors in statistics do not necessarily indicate a bug in the software, but rather a problem with the data or some interaction of the code and data. The statistician goes back and forth between cleaning, manipulating, modeling, visualizing and interpreting to study patterns and relations in the data. This simultaneous debugging of data and code comes down to a lot of trial and error. Problems with outliers, degrees of freedom or numeric properties do not reveal themselves until we try to fit a model or create a plot. Exceptions raised by statistical methods are often a sign that data needs additional work. This makes error messages an important source of information for the statistician to get to know a dataset and its intricacies. And while debugging the data we learn limitations of the analysis methods. In practice we sometimes find out that a particular dataset requires us to research or implement additional techniques because the standard tools do not suffice or are inappropriate.

Interactive error handling is one of the reasons that there is no clear distinction between development and production in scientific computing. When interfacing with analysis modules it is important that the role of errors is recognized. An \API must be able to handle exceptions and report error messages to the user, and certainly not crash the system. The role of errors and interactive debugging in data analysis can be confusing to developers outside of our community. Some popular commercial products seem to have propagated the belief that data analysis comes down to applying a magical formula to a dataset, and no intelligent action is required on the side of the user. Systems that only support such canned analyses don't do justice to the wide range of methods that statistics has to offer. In practice, interactive data debugging is an important concern of data analysis and embedded scientific computing.

\subsection{Security and resource control}

Somewhat related to the above are special needs in terms of security. Most statistical software currently available is primarily designed for interactive use on the local machine. Therefore access control is not considered an issue and the execution environment is entirely unrestricted. Embedded modules or public services require implementation of security policies to prevent malicious or excessive use of resources. This in itself is not a unique problem. Most scripting languages such as \texttt{php} or \texttt{python} do not enforce any access control and assume security will be implemented on the application level. But in the case of scientific computing, two domain specific aspects further complicate the problem. 

The first issue is that statistical software can be demanding and greedy with hardware resources. Numerical methods are expensive both in terms of memory and cpu. Fair-use policies are not really feasible because excessive use of resources often happens unintentionally. For example, an overly complex model specification or algorithm getting stuck could end up consuming all available memory and cpu until manually terminated. When this happens on the local machine, the user can easily interrupt the process prematurely by sending a \texttt{SIGINT} (pressing \texttt{CTRL+C} or \texttt{ESC}), but in a shared environment this needs to be regulated by the system. Embedded scientific computing requires technology and policies that can manage and limit memory allocation, cycles, disk space, concurrent processes, network traffic, etc. The degree of flexibility offered by implementation of resource management is an important factor in the scalability of a system. Fine grained control over system resources consumed by individual tasks allows for serving many users without sacrificing reliability. 

The second domain specific security issue is caused by the need for arbitrary code execution. A traditional application security model is based on user role privileges. In a standard web application, only a developer or administrator can implement and deploy actual code. The application merely exposes predefined functionality; users are not allowed to execute arbitrary code on the server. Any possibility of code injection is considered a security vulnerability and when found the server is potentially compromised. However as already mentioned, lack of segregation between users and developers in statistics gives limited use to applications that restrict users to predefined scripts and canned services. To support actual data analysis, the user needs access to the full language lexicon to freely explore and manipulate the data. The need for arbitrary code execution disqualifies user role based privileges and demands a more sophisticated security model. 
%Therefore the security policies are for the most part a concern of the server implementation and do not need to be described in the interface.

\subsection{Reproducible research}

Replication of findings is one of the main principles of the scientific method. In quantitative research, a necessary condition for replication is reproducibility of results. The goal of reproducible research is to tie specific instructions to data analysis and experimental data so that scholarship can be recreated, better understood, and verified \citep{cranRR}. Even though the ideas of replication are as old as science itself, reproducibility in scientific computing is still in its infancy. Tools are available that assist users in documenting their actions, but to most researchers these are not a natural part of the daily workflow. Fortunately, the importance of replication in data analysis is increasingly acknowledged and support for reproducibility is becoming more influential in the design of statistical software.

Reproducibility changes what constitutes the main product of data analysis. Rather than solely output and conclusions, we are interested recording and publishing the entire \emph{analysis process}. This includes all data, code and results but also external software that was used arrive at the results. Reproducibility puts high requirements on software versioning. More than in other fields it is crucial that we diligently archive and administer the precise versions or branches of all scripts, packages, libraries, plugins that were somehow involved in the process. If an analysis involves randomness, it is also important that we keep track of which seeds and random number generators were used. In the current design of statistical software, reproducibility was mostly an afterthought and has to be taken care of manually. In practice it is tedious and error-prone. There is a lot of room for improvement through software that incorporates reproducible practices as a natural part of the data analysis process.

Whereas reproducibility in statistics is acknowledged from a transparency and accountability point of view, it has enormous potential to become much more than that. There are interesting parallels between reproducible research and revision control in source code management systems. Technology for automatic reproducible data analysis could revolutionize scientific collaboration, similar to what \texttt{git} has done for software development. A system that keeps track of each step in the  analysis process like a \emph{commit} in software versioning would make peer review or follow-up analysis more practical and enjoyable. 
%A scientific publication would no longer be considered an end product, but rather the starting point of scientific debate. 
When colleagues or reviewers can easily reproduce results, test alternative hypotheses or recycle data, we achieve greater trustworthiness but also multiply return on investment of our work. Finally an open kitchen can help facilitate more natural ways of learning and teaching statistics. Rather than relying on general purpose textbooks with artificial examples, scholars directly study the practices of prominent researchers to understand how methods are applied in the context of data and problems as they appear specifically in their area of interest.

% !TEX root = main.tex
\section{The state problem}

Management of \emph{state} is a fundamental principle around which digital communications are designed. We distinguish \emph{stateful} and \emph{stateless} communication. In a stateless communication protocol, interaction involves independent request-response messages in which each request is unrelated by any previous request \citep{hennessy2012computer}. Because the messages are independent, there is no particular ordering to them and requests can be performed concurrently. Examples of stateless protocols include the internet protocol (\IP) and the hypertext transfer protocol (\HTTP). A stateful protocol on the other hand consists of an interaction via an ordered sequence of interrelated messages. The specification typically prescribes a specific mechanism for initiating and terminating a persistent \emph{connection} for information exchange. Examples of stateful protocols include the transmission control protocol (\TCP) or file transfer protocol (\texttt{FTP}).

%Different notions of state can exist in different layers of communications. For example \TCP provides stateful connections in a layer on top \texttt{IP}, by specifying procedures for labeling and ordering \IP messages. Conversely, \HTTP is a stateless application layer that builds on \TCP. When using \HTTP, requests are independent of each other and not part of a persistent connection. However, many websites that use \HTTP do require some notion of state, for example to distinguish between requests from various authenticated users. In such applications, the client typically includes a special session-key in the payload of each \HTTP request. This key a unique value that helps the server determine which requests were part of one and the same session which allows the server to retain the state of each such user session.

In most data analysis software, the user controls an interactive session through a console or \GUI, with the possibility of executing a sequence of operations in the form of a \emph{script}. Scripts are useful for publishing code, but the most powerful way of using the software is interactively. In this respect, statistical software is not unlike to a shell interface to the operating system. Interactivity in scientific computing makes management of state the most central challenge in the interface design. When moving from a \UI to \API perspective, support for statefulness becomes substantially more complicated. This section discusses how the existing bridges to \R have approached this problem, and their limitations. We then continue by explaining how the \OpenCPU \API exploits the functional paradigm to implement a hybrid solution that abstracts the notion of state and allows for a high degree of performance optimization.

\subsection{Stateless solutions: predefined scripts}

The easiest solution is to not incorporate state on the level of the interface, and limit the system to predefined scripts. This is the standard approach in traditional web development. The web server exposes a parameterized service which generates dynamic content by calling out to a script on the system via \CGI. Any support for state has to be implemented manually in the application layer, e.g. by writing code that stores values in a database. For \R we can use \texttt{rApache} \citep{horner2013rapache} to develop this kind of scripted applications very similar to web scripting languages such as \texttt{php}. This suffices for relatively simple services that expose limited, predefined functionality. Scripted solutions give the developer flexibility to freely define input and output that are needed for a particular application. For example, we can write a script that generates a plot based on a couple of input parameters and returns a fixed size \texttt{png} image. Because scripts are stateless, multiple requests can be performed concurrently. A lot of the early work in this research has been based on this approach, which is a nice starting point but becomes increasingly problematic for more sophisticated applications.

The main limitation of scripts is that to support basic interactivity, retention of state needs to be implemented manually in the application layer. A minimal application in statistics consists of the user uploading a data file, performing some manipulations and then creating a model, plot or report. When using scripts, the application developer needs to implement a framework to manage requests from various user sessions, and store intermediate results in a database or disk. Due to the complexity of objects and data in \R, this is much more involved than it is in e.g. \texttt{php}, and requires programming skills. Furthermore it leads to code that intermingles scientific computing with application logic, and rapidly increases complexity as the application gets extended with additional scripts.
%Storing data on disk can also introduce significant performance overhead. 
Because these problems will recur for almost any statistical application, we could benefit greatly from a system that supports retaining state by design.

Moreover predefined scripts are problematic because they divide developers and users in a way that is not very natural for scientific computing. Scripts in traditional web development give the client very little power to prevents malicious use of services. However, in scientific computing, a script often merely serves as a starting point for analysis. The user wants to be able to modify the script to look at the data in another way by trying additional methods or different procedures. A system that only allows for performing scripted actions severely handicaps the client and creates a lot of work for developers: because all functionality has to be prescripted, they are in charge of designing and implementing each possible action the user might want to perform. This is impractical for statistics because of the infinite amount of operations that can be performed on a dataset. For these reasons, the stateless scripting approach does not scale well to many users or complex applications.

\subsection{Stateful solution: client side process management}

Most existing bridges to \R have taken a stateful approach. Tools such as \texttt{Rserve} \citep{urbanek2013rserve} and \texttt{shiny} \citep{shiny} expose a low-level interface to a private \R process over a (web)socket. This gives clients freedom to run arbitrary \R code, which is great for implementing something like a web-based console or \texttt{IDE}. The main problem with existing stateful solutions is lack of interoperability. Because these tools are in essence a remote \R console, they do not specify any standardized interface for calling methods, data \texttt{I/O}, etc. A low-level interface requires extensive knowledge of logic and internals of \R to communicate, which again leads to high coupling. The client needs to be aware of \R syntax to call \R methods, interpret \R data structures, capture graphics, etc. These bridges are typically intended to be used in combination with a special client. In the case of  \texttt{shiny}, the server comes with a set of widget templates that can be customized from within \R. This allows \R users to create a basic web \GUI without writing any \texttt{HTML} or \JavaScript, which can be very useful. However, the shiny software is not designed for integration with non-shiny clients and serves a somewhat different purpose and audience than tools for embedded scientific computing.

Besides high coupling and lack of interoperability, stateful bridges also introduce some technical difficulties. Systems that allocate a private \R process for each client cannot support concurrent requests within a session. Each incoming request has to wait until the previous requests are finished for the process to become available. In addition to suboptimal performance, this can also be a source of instability. When the \R process gets stuck or raises an unexpected error, the server might become unresponsive causing the application to crash. Another drawback is that stateful servers are extremely expensive and inefficient in terms of memory allocation. The server has to keep each \R process alive for the full duration of a session, even when idle most of the time. Memory that is in use by any single client does not free up until the user closes the application. This is particularly unfortunate because memory is usually the main bottleneck in data intensive applications of scientific computing. Moreover, connectivity problems or ill-behaved clients require mechanisms to timeout and terminate inactive processes, or save and restore an entire session.

\subsection{A hybrid solution: functional state}

We can take the best of both worlds by abstracting the notion of state to a higher level. Interactivity and state in \OpenCPU is provided through persistence of \emph{objects} rather than a persistent \emph{process}. As it turns out, this is a natural and powerful definition of state within the functional paradigm. Functional programming emphasizes that output from methods depends only on their inputs and not on the program state. Therefore, functional languages can support state without keeping an entire process alive: merely retaining the state of objects should be sufficient. As was discussed before, this has obvious parallels with mathematics, but also maps beautifully to stateless protocols such as \HTTP. The notion of state as the set of objects is already quite natural to the \R user, as is apparent from the \texttt{save.image} function. This function serializes all objects in the global environment to a file on disk which described in the documentation as ``saving the current workspace''. Exploiting this same notion of state in our interface allows us to get the benefits of both traditional stateless and stateful approaches without introducing additional complexity. This simple observation provides the basis for a very flexible, stateful \RPC system.

%to colloquial
To facilitate this, the \OpenCPU \API defines a mapping between \HTTP requests and \R function calls. After executing a function call, the server stores all outputs (return value, graphics, files) and a \emph{temporary key} is given to the client. This key can be used to control these newly created resources in future requests. The client can retrieve objects and graphics in various formats, publish resources, or use them as arguments in subsequent function calls. An interactive application consists of a series of \RPC requests with keys referencing the objects to be reused as arguments in consecutive function calls, making the individual requests technically stateless. Besides reduced complexity, this system makes parallel computing and asynchronous requests a natural part of the interaction. To compute $f(g(x), h(y))$, the client can perform \RPC requests for $g(x)$ and $h(y)$ simultaneously and pass the resulting keys to $f()$ in a second step. In an asynchronous client language such as \texttt{JavaScript} this happens so naturally that it requires almost no effort from the user or application developer.

One important detail is that \OpenCPU deliberately does not prescribe how the server should implement storing and loading of objects in between requests. The \API only specifies a system for performing \R function calls over \HTTP and referencing objects from keys. Different server implementations can use different strategies for retaining such objects. A naive implementation could simply serialize objects to disk after each request and immediately terminate the process. This is safe and easy, but writing to disk can be slow. A more sophisticated implementation could keep objects in memory for a while longer, either by keeping the \R process alive or through some sort of in-memory database or memcached system. 
Thereby the resources do not need to be loaded from disk if they are reused in a subsequent request shortly after being created. This illustrates the kind of optimization that can be achieved by carefully decoupling server and client components.

% !TEX root = main.tex
\section{The OpenCPU HTTP API}

This section introduces the most important concepts and operations of the \API. At this point the concerns discussed in earlier chapters become more concrete as we illustrate how the pieces come together in the context of \R and \HTTP. It is not the intention to provide a detailed specification of every feature of the system. We focus on the main parts of the interface that exemplify the separation of concerns central to this work. The online documentation and reference implementations are the best source of information on the specifics of implementing clients and applications.

\subsection{About HTTP}

One of the major strengths of \OpenCPU is that it builds on the hypertext transfer protocol \citep{rfc2616}. \HTTP is the most used application protocol on the internet, and the foundation of data communication in browsers and the world wide web. The \HTTP specification is very mature and widely implemented. It provides all functionality required to build modern applications and has recently gained popularity for web \API's as well. The benefit of using a standardized application protocol is that a lot of functionality gets built-in by design. \HTTP has excellent mechanisms for authentication, encryption, caching, distribution, concurrency, error handling, etc. This allows us to defer most application logic of our system to the protocol and limit the \API specification to logic of scientific computing. 

The \OpenCPU \API defines a mapping between \HTTP requests and high-level operations such as calling functions, running scripts, access to data, manual pages and management of files and objects. The \API deliberately does not prescribe any language implementation details. Syntax and low-level concerns such as process management or code evaluation are abstracted and at the discretion of the server implementation. The \API also does not describe any logic which can be taken care of on the protocol or application layer. For example, to add support for authentication, any of the standard mechanisms can be used such as \texttt{basic auth} \citep{franks1999rfc} or \texttt{OAuth 2.0} \citep{hardt2012oauth}. The implementation of such authentication methods might vary from a simple server configuration to defining additional endpoints. But because authentication will not affect the meaning of the \API itself, it can be considered independent of this research. The same holds for other features of the \HTTP protocol which can be used in conjunction with the \OpenCPU \API (or any other \HTTP interface for that matter). What remains after cutting out implementation and application logic is a simple and interoperable interface that is easy to understand and can be implemented with standard \HTTP software libraries. This is an enormous advantage over many other bridges to \R and critical to make the system scalable and extensible. 

% !TEX root = main.tex
\subsection{Resource types}

As was described earlier, individual requests within the \OpenCPU \API are stateless and there is no notion of a \emph{process}. State of the system changes through creation and manipulation of resources. This makes the various resource types the conceptual building blocks of the \API. Each resource type has unique properties and supports different operations.

\subsubsection{Objects}

Objects are the main entities of the system and carry the same meaning as within a functional language. They include data structures, functions, or other types supported by the back-end language, in this case \R. Each object has an individual endpoint within the \API and unique name or key within its namespace. The client needs no knowledge of the implementation of these objects. Analogous to a \UI, the primary purpose of the \API is managing objects (creating, retrieving, publishing) and performing procedure calls. Objects created from executing a script or returned by a function call are automatically stored and gain the same status as other existing objects. The \API does not distinguish between static objects that appear in e.g. packages, or dynamic objects created by users, nor does it distinguish between objects in memory or on disk. The \API merely provides a system for referencing objects in a way that allows clients to control and reuse them. The implementation of persistence, caching and expiration of objects is at the discretion of the server. 

\subsubsection{Namespaces}

A namespace is a collection of uniquely named objects with a given path in the \API. In \R, static namespaces are implemented using \emph{packages} and dynamic namespaces exist in \emph{environments} such as the user workspace. \OpenCPU abstracts the concept of a namespace as a set of uniquely named objects and does not distinguish between static, dynamic, persistent or temporary namespaces. Clients can request a list of the contents of any namespace, yet the server might refuse such a request for private namespaces or hidden objects. 

\subsubsection{Formats}

\OpenCPU explicitly differentiates a resource from a \emph{representation} of that resource in a particular \emph{format}. The \API lets the client rather than the server decide on the format used to serve content. This is a difference with common scientific practices of exchanging data, documents and figures in fixed format files. Resources in \OpenCPU can be retrieved using various output formats and formatting parameters. For example, a basic dataset can be retrieved in \texttt{csv}, \texttt{json}, \texttt{Protocol Buffers} or \texttt{tab delimited} format. Similarly, a graphic can be retrieved in \texttt{svg}, \texttt{png} or \texttt{pdf} and manual pages can be retrieved in \texttt{text}, \texttt{html} or \texttt{pdf} format. In addition to the format, the client can specify formatting parameters in the request. The system supports many additional formats, but not every format is appropriate for every resource type. When a client requests a resource in a format using an invalid format, the server responds with an error.

\subsubsection{Data}

The \API defines a separate entity for \emph{data} objects. Even though data can technically be treated as general objects, they often serve a different purpose. Data are usually not language specific and cannot be called or executed. Therefore it can be useful to conceptually distinguish this subclass. For example, \R uses lazy loading of data objects to save memory when for packages containing large datasets.

\subsubsection{Graphics}

Any function call can produce zero or more graphics. After completing a remote function call, the server reports how many graphics were created and provides the key for referencing these graphics. Clients can retrieve each individual graphic in subsequent requests using one of various output formats such as \texttt{png}, \texttt{pdf}, and \texttt{svg}. Where appropriate the client can specify additional formatting parameters during the retrieval of the graphic such as width, height or font size.

\subsubsection{Files}

Files can be uploaded and downloaded using standard \HTTP mechanics. The client can post a file as an argument in a remote function call, or download files that were saved to the working directory by the function call. Support for files also allows for hosting web pages (e.g. \texttt{html}, \texttt{css}, \texttt{js}) that interact with local \API endpoints to serve a web application. Furthermore files that are recognized as \emph{scripts} can be executed using \RPC.

\subsubsection{Manuals}

In most scientific computing languages, each function or dataset that is available to the user is accompanied by an identically named manual page. This manual page includes information such as description and usage of functions and their arguments, or comments about the columns of a particular dataset. Manual pages can be retrieved through the \API in various formats including \texttt{text}, \texttt{html} and \texttt{pdf}.

\subsubsection{Sources}

The \OpenCPU specification makes reproducibility an integrated part of the \API interaction. In addition to results, the server stores the call and arguments for each \RPC request. The same key that is used to retrieve objects or graphics can be used to retrieve sources or automatically replicate the computation. Hence for each output resource on the system, clients can lookup the code, data, warnings and packages that were involved in its creation. Thereby results can easily be recalculated, which forms a powerful basis for reproducible practices. This feature can be used for other purposes as well. For example, if a function fetches dynamic data from an external resource to generate a model or plot, reproduction is used to \emph{update} the model or plot with new data.

\subsubsection{Containers}

We refer to a path on the server containing one or more collections of resources as a \emph{container}. The current version of \OpenCPU implements two types of containers. A \emph{package} is a static container which may include a namespace with \R objects, manual pages, data and files.  A \emph{session} is a dynamic container which holds outputs created from executing a script or function call, including a namespace with \R objects, graphics and files. The distinction between packages and sessions is an implementation detail. The \API does not differentiate between the various container types: interacting with an object or file works the same, regardless of whether it is part of a package or session. Future versions or other servers might implement different container types for grouping collections of resources.

\subsubsection{Libraries}

We refer to a collection of containers as a \emph{library}. In \R terminology, a library is a directory on disk with installed packages. Within the context of the \API, the concept is not limited to packages but refers more generally to any set of containers. The \texttt{/ocpu/tmp/} library for example is the collection of temporary sessions. Also the \API notion of a library does not require containers to be preinstalled. A remote collection of packages, which in \R terminology is called a \emph{repository}, can also be implemented as a library. The current implementation of \OpenCPU exposes the \texttt{/ocpu/cran/} library which refers to the current packages on the \texttt{CRAN} repository. The \API does not differentiate between a library of sessions, local packages or remote packages. Interacting with an object from a \texttt{CRAN} package works the same as interacting with an object from a local package or temporary session. The \API leaves it up to the server which types of libraries it wishes to expose and how to implement this. The current version of \OpenCPU uses a combination of cron-jobs and on-the-fly package installations to synchronize packages on the server with the \texttt{CRAN} repositories.

\subsection{Methods}

The current \API uses two \HTTP methods: \GET and \POST. As per \HTTP standards, \GET is a \emph{safe} method which means it is intended only for information reading and should not change the state of the server. \OpenCPU uses the \GET method to retrieve objects, manuals, graphics or files. The parameters of the request are mapped to the formatting function. A \GET requests targeting a container, namespace or directory is used to list the contents. The \POST method on the other is used for \RPC which does change server state. A \POST request targeting a function results in a remote function call where the \HTTP parameters are mapped to function arguments. A \POST request targeting a script results in an execution of the script where \HTTP parameters are mapped to the script interpreter. Table \ref{table:methods} gives an overview using the \texttt{MASS} package \citep{MASS} as an example.

\begin{table}[H]
\centering
\def\arraystretch{1.3}%
\begin{tabular}{@{}lllll@{}}
\toprule
\emph{Method} & \emph{Target} & \emph{Action}  & \emph{Parameters}     & \emph{Example}                                      \\ \midrule
\texttt{GET}    & object  & retrieve      &  formatting     & \texttt{GET /ocpu/library/MASS/data/cats/json}            \\
                & manual  & read          &  formatting     & \texttt{GET /ocpu/library/MASS/man/rlm/html}            \\  
                & graphic & render        &  formatting    & \texttt{GET /ocpu/tmp/\{key\}/graphics/1/png}            \\   
                & file    & download      & -                     & \texttt{GET /ocpu/library/MASS/NEWS}                         \\
                & path    & list contents & -                     & \texttt{GET /ocpu/library/MASS/scripts/}                     \\ \midrule
\texttt{POST}   & object  & call function & function arguments    & \texttt{POST /ocpu/library/stats/R/rnorm}                    \\
                & file    & run script    & control interpreter   & \texttt{POST /ocpu/library/MASS/scripts/ch01.R}              \\ \bottomrule
\end{tabular}
\caption{Currently implemented \HTTP methods}
\label{table:methods}
\end{table}

\subsection{Status codes}

Each \HTTP response includes a status code. Table \ref{table:statuscodes} lists some common \HTTP status codes used by \OpenCPU that the client should be able to interpret. The meaning of these status codes is conform \HTTP standards. The web server may use additional status codes for more general purposes that are not specific to \OpenCPU.

\begin{table}[H]
\centering
\def\arraystretch{1.3}%
%\noindent\begin{tabular}{\columnwidth}{ *{3}{X} }
\begin{tabular}{@{}lll@{}}
\toprule
\emph{Status Code}              & \emph{Happens when}                             & \emph{Response content}                     \\ \midrule
\texttt{200 OK}          & On successful \texttt{GET} request                     & Requested data                    \\
\texttt{201 Created}     & On successful \texttt{POST} request                    & Output key and location                     \\
\texttt{302 Found}       & Redirect                                               & Redirect location                   \\
\texttt{400 Bad Request} & On computational error in \R                                     & Error message from \R in \texttt{text/plain} \\
\texttt{502 Bad Gateway} & Back-end server offline                            & -- (See error logs) \\
\texttt{503 Bad Request} & Back-end server failure                                & -- (See error logs) \\ \bottomrule                          
\end{tabular}
\caption{Commonly used \HTTP status codes}
\label{table:statuscodes}
\end{table}

\subsection{Content-types}

Clients can retrieve objects in various \emph{formats} by adding a format identifier suffix to the \URL in a \GET request. Which formats are supported and how object types map to a particular format is at the discretion of the server implementation. Not every format can support any object type. For example, \texttt{csv} can only be used to retrieve tabular data structures and \texttt{png} is only appropriate for graphics. Table \ref{table:formats} lists the formats \OpenCPU supports, the respective internet media type, and the \R function that \OpenCPU uses to export an object into a particular format. Arguments of the \GET requests are mapped to this export function. The \texttt{png} format has parameters such as \texttt{width} and \texttt{height} as documented in \texttt{?png}, whereas the \texttt{tab} format has parameters \texttt{sep}, \texttt{eol}, \texttt{dec} which specify the delimiting, end-of-line and decimal character respectively as documented in \texttt{?write.table}.

\begin{table}[H]
\centering
\def\arraystretch{1.3}%
\begin{tabular}{@{}lllll@{}}
\toprule
 \emph{Format} & \emph{Content-type}             & \emph{Export function}      & \emph{Example}    \\ \midrule
 \texttt{print}  & \texttt{text/plain}               & \texttt{base::print}    & \texttt{/ocpu/cran/MASS/R/rlm/print}          \\
 \texttt{rda}    & \texttt{application/octet-stream} & \texttt{base::save}     & \texttt{/ocpu/cran/MASS/data/cats/rda}          \\
 \texttt{rds}    & \texttt{application/octet-stream} & \texttt{base::saveRDS}  & \texttt{/ocpu/cran/MASS/data/cats/rds}          \\
 \texttt{json}   & \texttt{application/json}         & \texttt{jsonlite::toJSON}   & \texttt{/ocpu/cran/MASS/data/cats/json}      \\
 \texttt{pb}     & \texttt{application/x-protobuf}   & \texttt{RProtoBuf::serialize\_pb} & \texttt{/ocpu/cran/MASS/data/cats/pb} \\
 \texttt{tab}    & \texttt{text/plain}               & \texttt{utils::write.table}   & \texttt{/ocpu/cran/MASS/data/cats/tab}    \\
 \texttt{csv}    & \texttt{text/csv}                 & \texttt{utils::write.csv}    & \texttt{/ocpu/cran/MASS/data/cats/csv}     \\
 \texttt{png}    & \texttt{image/png}                & \texttt{grDevices::png}      & \texttt{/ocpu/tmp/\{key\}/graphics/1/png}    \\
 \texttt{pdf}    & \texttt{application/pdf}          & \texttt{grDevices::pdf}      & \texttt{/ocpu/tmp/\{key\}/graphics/1/pdf}     \\
 \texttt{svg}    & \texttt{image/svg+xml}            & \texttt{grDevices::svg}      & \texttt{/ocpu/tmp/\{key\}/graphics/1/svg}     \\ \bottomrule
\end{tabular}
\caption{Currently supported export formats and corresponding \texttt{Content-type}}
\label{table:formats}
\end{table}

\subsection{URLs}

The root of the \API is dynamic, but defaults to \texttt{/ocpu/} in the current implementation. Clients should make the \OpenCPU server address and root path configurable. In the examples we assume the defaults. As discussed before, \OpenCPU currently implements two container types to hold resources. Table \ref{table:packageapi} lists the \texttt{URL}s of the \emph{package} container type, which includes objects, data, manual pages and files.

\begin{table}[H]
\centering
\def\arraystretch{1.3}%
\begin{tabular}{@{}lll@{}}
\toprule
\emph{Path} & \emph{Description}                      & \emph{Examples}                \\ \midrule
\texttt{.}    & Package information                      & \texttt{/ocpu/cran/MASS/}               \\
\texttt{./R}    & Exported namespace objects             & \texttt{/ocpu/cran/MASS/R/}             \\
     &                                                   & \texttt{/ocpu/cran/MASS/R/rlm/print}    \\
\texttt{./data} & Data objects in the package (\HTTP \GET only)         & \texttt{/ocpu/cran/MASS/data/}          \\
     &                                                   & \texttt{/ocpu/cran/MASS/data/cats/json} \\
\texttt{./man}  & Manual pages in the package (\HTTP \GET only)         & \texttt{/ocpu/cran/MASS/man/}           \\
     &                                                   & \texttt{/ocpu/cran/MASS/man/rlm/html}   \\
\texttt{./*}    & Files in installation directory, relative to package the root      & \texttt{/ocpu/cran/MASS/NEWS}    \\
     &                                                   & \texttt{/ocpu/cran/MASS/scripts/}       \\ \bottomrule
\end{tabular}
\caption{The package container includes objects, data, manual pages and files.}
\label{table:packageapi}
\end{table}

Table \ref{table:sessionapi} lists \texttt{URL}s of the \emph{session} container type. This container holds outputs generated from a \RPC request and includes objects, graphics, source code, stdout and files. As noted earlier, the distinction between packages and sessions is considered an implementation detail. The \API does not differentiate between objects and files that appear in packages or in sessions.

\begin{table}[H]
\centering
\def\arraystretch{1.3}%
\begin{tabular}{@{}lll@{}}
\toprule
\emph{Path}          & \emph{Description}                      & \emph{Examples}                \\ 
\midrule
\texttt{.}          & Session content list                                       & \texttt{/ocpu/tmp/\{key\}/}               \\
\texttt{./R}        & Objects created by the \RPC request                        & \texttt{/ocpu/tmp/\{key\}/R/}             \\
                    &                                                            & \texttt{/ocpu/tmp/\{key\}/R/mydata/json}  \\
\texttt{./graphics} & Graphics created by the \RPC request                       & \texttt{/ocpu/tmp/\{key\}/graphics/}      \\
                    &                                                            & \texttt{/ocpu/tmp/\{key\}/graphics/1/png} \\
\texttt{./source}   & Source code of \RPC request                                & \texttt{/ocpu/tmp/\{key\}/source}       \\
\texttt{./stdout}   & \texttt{STDOUT} from by the \RPC request                   & \texttt{/ocpu/tmp/\{key\}/stdout}       \\
\texttt{./console}  & Mixed source and \texttt{STDOUT} emulating console output  & \texttt{/ocpu/tmp/\{key\}/console}     \\
\texttt{./files/*}  & Files saved to working dir by the \RPC request             & \texttt{/ocpu/tmp/\{key\}/files/myfile.xyz}       \\

\bottomrule
\end{tabular}
\caption{The session container includes objects, graphics, source, stdout and files.}
\label{table:sessionapi}
\end{table}

\subsection{RPC requests}

A \POST request in \OpenCPU always invokes a remote procedure call (\RPC). Requests targeting a \emph{function} object result in a function call where the \HTTP parameters from the post body are mapped to function \emph{arguments}. A \texttt{POST} targeting a \emph{script} results in execution of the script where \HTTP parameters are passed to the script interpreter. The term \RPC refers to both remote function calls and remote script executions. The current \OpenCPU implementation recognizes scripts by their file extension, and supports \R, \texttt{latex}, \texttt{markdown}, \texttt{Sweave} and \texttt{knitr} scripts. Table \ref{table:scripts} lists each script type with the respective file extension and interpreter.

\begin{table}[H]
\centering
\def\arraystretch{1.3}%
\begin{tabular}{@{}lll@{}}
\toprule
\emph{File extension} & \emph{Type}           & \emph{Interpreter}                   \\ \midrule
\texttt{file.r}         & \R       & \texttt{evaluate::evaluate}            \\
\texttt{file.tex}       & \Latex          & \texttt{tools::texi2pdf}               \\
\texttt{file.rnw}       & \texttt{knitr}/\texttt{sweave}   & \texttt{knitr::knit} + \texttt{tools::texi2pdf} \\
\texttt{file.md}        & \texttt{markdown}       & \texttt{knitr::pandoc}                 \\
\texttt{file.rmd}       & \texttt{knitr markdown} & \texttt{knitr::knit} + \texttt{knitr::pandoc}   \\
\texttt{file.brew}      & \texttt{brew}           & \texttt{brew::brew}                    \\ \bottomrule
\end{tabular}
\caption{Files recognized as scripts and their characterizing file extension}
\label{table:scripts}
\end{table}

An important conceptual difference with a terminal interface is that in the \OpenCPU \API, the server determines the namespace that output of a function call is assigned to. The server includes a \emph{temporary key} in the \RPC response that serves the same role as a variable name. The key and is used to reference the newly created resources in future requests. Besides the return value, the server also stores graphics, files, warnings, messages and \texttt{stdout} that were created by the \RPC. These can be listed and retrieved using the same key. In \R, the function call itself is also an object which is added to the collection for reproducibility purposes. 

Objects on the system are non-mutable and therefore the client cannot change or overwrite existing keys. For functions that modify the state of an object, the server creates a copy of the modified resource with a new key and leaves the original unaffected.

\subsection{Arguments}

Arguments to a remote function call can be posted using one of several methods. A data interchange format such as \JSON or \texttt{Protocol Buffers} can be used to directly post data structures such as lists, vectors, matrices or data frames. Alternatively the client can reference the name or key of an existing object. The server automatically resolves keys and converts interchange formats into objects to be used as arguments in the function call. Files contained in a \texttt{multipart/form-data} payload of an \RPC request are copied to the working directory and the argument of the function call is set to the filename. Thereby, remote function calls with a file arguments can be performed using standard \HTML form submission.

\begin{table}[H]
\centering
\def\arraystretch{1.3}%
\begin{tabular}{@{}llllll@{}}
\toprule
\emph{Content-type}                      & \emph{Primitives} & \emph{Data structures}  &  \emph{Raw code} & \emph{Files} & \emph{Temp key} \\ \midrule
\texttt{multipart/form-data}               & OK         & OK (inline \texttt{json}) & OK       & OK   & OK            \\
\texttt{application/x-www-form-urlencoded} & OK         & OK (inline \texttt{json}) & OK       & -    & OK            \\
\texttt{application/json}                  & OK         & OK               & -        & -    & -             \\
\texttt{application/x-protobuf}            & OK         & OK               & -        & -    & -             \\ \bottomrule
\end{tabular}
\caption{Accepted request \texttt{Content-types} and supported argument formats}
\label{table:arguments}
\end{table}

The current implementation supports several standard \texttt{Content-type} formats for passing arguments to a remote function call within a \POST request, including \texttt{application/x-www-form-urlencoded}, \texttt{multipart/form-data}, \texttt{application/json} and \texttt{application/x-protobuf}. Each parameter or top level field within a \POST payload contains a single argument value. Table \ref{table:arguments} shows a matrix supported argument formats for each \texttt{Content-types}.

\subsection{Privacy}

Because the data and sources of a statistical analysis include potentially sensitive information, the temporary keys from \RPC requests are private. Clients should default to keeping these keys secret, given that leaking a key will compromise confidentiality of their data. The system does not allow clients to search for keys or retrieve resources without providing the appropriate key. In this sense, a temporary key has a similar status as an \emph{access token}. Because temporary keys are private, multiple users can share a single \OpenCPU server without any form of authentication. Each request is anonymous and confidential, and only the client that performed the \RPC has the key to access resources from a particular request.

However, temporary keys do not have to be kept private per se: clients can choose to exchange keys with other clients. Unlike typical access tokens, the keys in \OpenCPU are unique for each request. Hence by publishing a particular key, the client reveals only the resources from a specific \RPC request, and no other confidential information. Resources in \OpenCPU are not tied to any particular user, in fact, there are no users in \OpenCPU system itself. Clients can share objects, graphics or files with each other, simply by communicating keys to these resources. Because each key holds both the output as well as the sources for an \RPC request, shared objects are reusable and reproducible by design. In some sense, all clients share a single universal namespace with keys containing hidden objects from all \RPC requests. By knowing a key to a particular resource it can be used as any other object on the system. This shapes the contours of a social analysis platform in which users collaborate by sharing reproducible, reusable resources identified by unique keys.